\documentclass[10pt]{article}
\usepackage[T1]{fontenc}
\usepackage[font=small]{caption}
\usepackage{graphicx,verbatim}
\usepackage{bbding}
\usepackage{hyperref}
\usepackage{color}

\usepackage{float}
\usepackage{amsmath}
\usepackage{amsfonts}
\usepackage{cite}  
\usepackage{natbib}

\usepackage{authblk}

\begin{document}
\title{\Large Learning Mechanistic Subtypes of Neurodegeneration with a Physics-Informed Variational Autoencoder Mixture Model}
\author[1]{Sanduni Pinnawala\thanks{m.pinnawala@sussex.ac.uk}}
\author[1,2]{Annabelle Hartanto} 
\author[1]{Ivor J. A. Simpson}
\author[1]{Peter A. Wijeratne}
\affil[1]{Sussex AI Centre, School of Engineering and Informatics, University of Sussex, UK}
\affil[2]{Sussex Neuroscience, School of Life Sciences, University of Sussex, UK} 

\date{\vspace{-1.25cm}}    
\maketitle 
\begin{abstract}
\small
Modelling the underlying mechanisms of neurodegenerative diseases demands methods that capture heterogeneous and spatially varying dynamics from sparse, high-dimensional neuroimaging data. Integrating partial differential equation (PDE) based physics knowledge with machine learning provides enhanced interpretability and utility over classic numerical methods. However, current physics-integrated machine learning methods are limited to considering a single PDE, severely limiting their application to diseases where multiple mechanisms are responsible for different groups (i.e., subtypes) and aggravating problems with model misspecification and degeneracy. Here, we present a deep generative model for learning mixtures of latent dynamic models governed by physics-based PDEs, going beyond traditional approaches that assume a single PDE structure. Our method integrates reaction-diffusion PDEs within a variational autoencoder (VAE) mixture model framework, supporting inference of subtypes of interpretable latent variables (e.g. diffusivity and reaction rates) from neuroimaging data. We evaluate our method on synthetic benchmarks and demonstrate its potential for uncovering mechanistic subtypes of Alzheimer's disease progression from positron emission tomography (PET) data.
\end{abstract}
\section{\large Introduction}
\normalsize
Neurodegenerative diseases, such as Alzheimer's disease (AD), are currently hypothesised to be caused by the spread of pathological proteins (tau and amyloid) across the brain \cite{buscheSynergyAmyloidvTau2020}. Positron emission tomography (PET) imaging is routinely used to obtain surrogate measures of tau and amyloid concentration in vivo; recently, longitudinal PET studies such as the Alzheimer's Disease Neuroimaging Initiative (ADNI) \cite{petersenAlzheimerDiseaseNeuroimaging2010} have provided measurements of tau and amyloid spatiotemporal dynamics, supporting the development of mechanistic models of disease propagation \cite{youngDatadrivenModellingNeurodegenerative2024}. The spatiotemporal dynamics are commonly modelled mathematically using partial differential equations (PDEs), which confer interpretability of model parameters and functional forms. Recently, PDEs have been integrated into machine learning frameworks to leverage their ability to scale inference to large, high dimensional datasets \cite{karniadakisPhysicsinformedMachineLearning2021}.

A challenge with using PDE-based approaches is that the structure and parameters of the PDEs underlying tau and amyloid propagation are not fully known. Further complicating the problem, neurodegenerative diseases such as AD are highly heterogeneous in progression between individuals, with disease clusters (i.e., subtypes) reported in both tau and amyloid PET \cite{collijDatadrivenEvidenceThree2021,vogelFourDistinctTrajectories2021}. Furthermore, the inverse problems of inferring these PDEs are fundamentally ill-posed, leading to several fundamental issues, including solution non-uniqueness \cite{arridgeSolvingInverseProblems2019}. Among the inference challenges, model misspecification -- where the modelled PDE does not reflect the true underlying PDE generating the data -- and model degeneracy -- where multiple distinct parameter sets or model structures yield indistinguishable outputs -- are inherent challenges when attempting to learn biophysical mechanistic models, further exacerbating the problem. These issues motivate an approach that can account for variability in both data and model structure to help support reliable inference.

\subsection{\normalsize Contributions}
\normalsize
We propose ``BrainPhys'', a physics-informed variational autoencoder (VAE) mixture model of neurodegenerative disease dynamics. Unlike approaches that assume a single governing PDE, we permit multiple possible PDEs simultaneously via mixture modelling, enabling our model to uncover subtypes of disease mechanisms within the same population. Our primary contributions are as follows.
\begin{enumerate}
\item We introduce a novel physics-guided VAE mixture model that can learn the parameters of multiple PDEs within the same dataset and be used to study issues regarding mechanistic model misspecification and degeneracy.
\item We demonstrate our model's ability to recover clusters of mechanistic models and their parameters using synthetic data.
\item We apply our model to a combined tau and amyloid PET dataset from ADNI, and find evidence supporting a 2-component mixture model.
\end{enumerate}

\subsection{\normalsize Related Work}
\normalsize
\textbf{Physics-Informed Modelling of Neurodegeneration.} Modelling the progression of neurodegenerative diseases has increasingly relied on physical models of protein propagation (e.g., prion-like spread) \cite{weickenmeierMultiphysicsPrionlikeDiseases2018, garbarinoModelingInferenceSpatiotemporal2019, iturria-medinaEpidemicSpreadingModel2014}. One influential line of work includes \cite{garbarinoInvestigatingHypothesesNeurodegeneration2021}, where the authors employed Gaussian processes to model and simulate the dynamics of protein concentration in the brain. However, these models often assume fixed dynamics and rely on strong priors, whereas our method enables more flexible latent modelling with physical constraints.

\textbf{Probabilistic Models for PDE-Constrained Systems.} Recent years have seen a surge in probabilistic approaches for learning solutions to parametric PDEs. These include Bayesian physics-informed neural networks \cite{yangBPINNsBayesianPhysicsInformed2021}, Gaussian processes with finite element methods \cite{duffinStatisticalFiniteElements2021}, normalising flow models \cite{guoNormalizingFieldFlows2022}, physics-driven deep latent variable models \cite{vadeboncoeurFullyProbabilisticDeep2023}. Notably, the physics-integrated VAE framework introduced in \cite{takeishiPhysicsIntegratedVariationalAutoencoders2021} incorporates physics constraints within the VAE architecture. Our approach builds upon this concept and extends to model structural variability in PDE dynamics. Instead of embedding all sources of uncertainty within a single neural surrogate, we adopt a mixture-model formulation that enables the latent space to reflect distinct PDE regimes (e.g., different reaction dynamics), enabling interpretable modelling of structural variability.

\textbf{Interpretable and Structured Latent Representations.} There is growing interest in learning interpretable latent spaces that correspond to physically meaningful variables. Some approaches focus on disentangled representations \cite{kimDisentanglingFactorising2019, chenIsolatingSourcesDisentanglement2019}, while others embed neural operators or physics-informed neural networks (PINNs) within generative models \cite{karniadakisPhysicsinformedMachineLearning2021, toscanoPINNsPIKANsRecent2024}. Our work contributes to this area by explicitly designing the latent space to represent a mixture of biophysically meaningful quantities (e.g., diffusion and reaction coefficients).

\section{\large Methodology}
\subsection{\normalsize Problem Setting}
\normalsize
We consider parameter learning in reaction-diffusion type equations of the form:
\begin{equation}
\frac{\partial}{\partial t}u (\textbf{x},t) = D \nabla^2 u(\textbf{x},t) + f(u(\textbf{x},t)),   \text{ on } \Omega, 
\label{eq1} 
\end{equation}
\begin{equation}
\nabla u(\textbf{x},t) \cdot \textbf{n} = 0, \text{ on } \partial\Omega,
\label{eq2} 
\end{equation}
\begin{equation}
u(\textbf{x},t) = u_0(\textbf{x}), \text{ for t = 0}.
\label{eq3} 
\end{equation}
Here $u(\textbf{x},t)$ is a scalar-valued function of space, $\textbf{x}$, and time, $t$; $D$ is the diffusion coefficient; and $f(\textbf{x},t)$ is the reaction term; $\Omega$ and $\partial\Omega$ are the domain and boundary, respectively; $\textbf{n}$ is the normal unit vector on the surface $\partial\Omega$; and $u_0(\textbf{x})$ is the initial field at the start of the simulation. Models of this type are used extensively for modelling biophysical processes, e.g., \cite{thompsonPrimerReactionDiffusion2018, erbanStochasticModellingReaction2009, kondoReactionDiffusionModelFramework2010}.

\subsection{\normalsize BrainPhys: A VAE Mixture of Physics Models}
\label{sec:2.2}
\begin{figure}[h!]
\centering
\includegraphics[width=\textwidth]{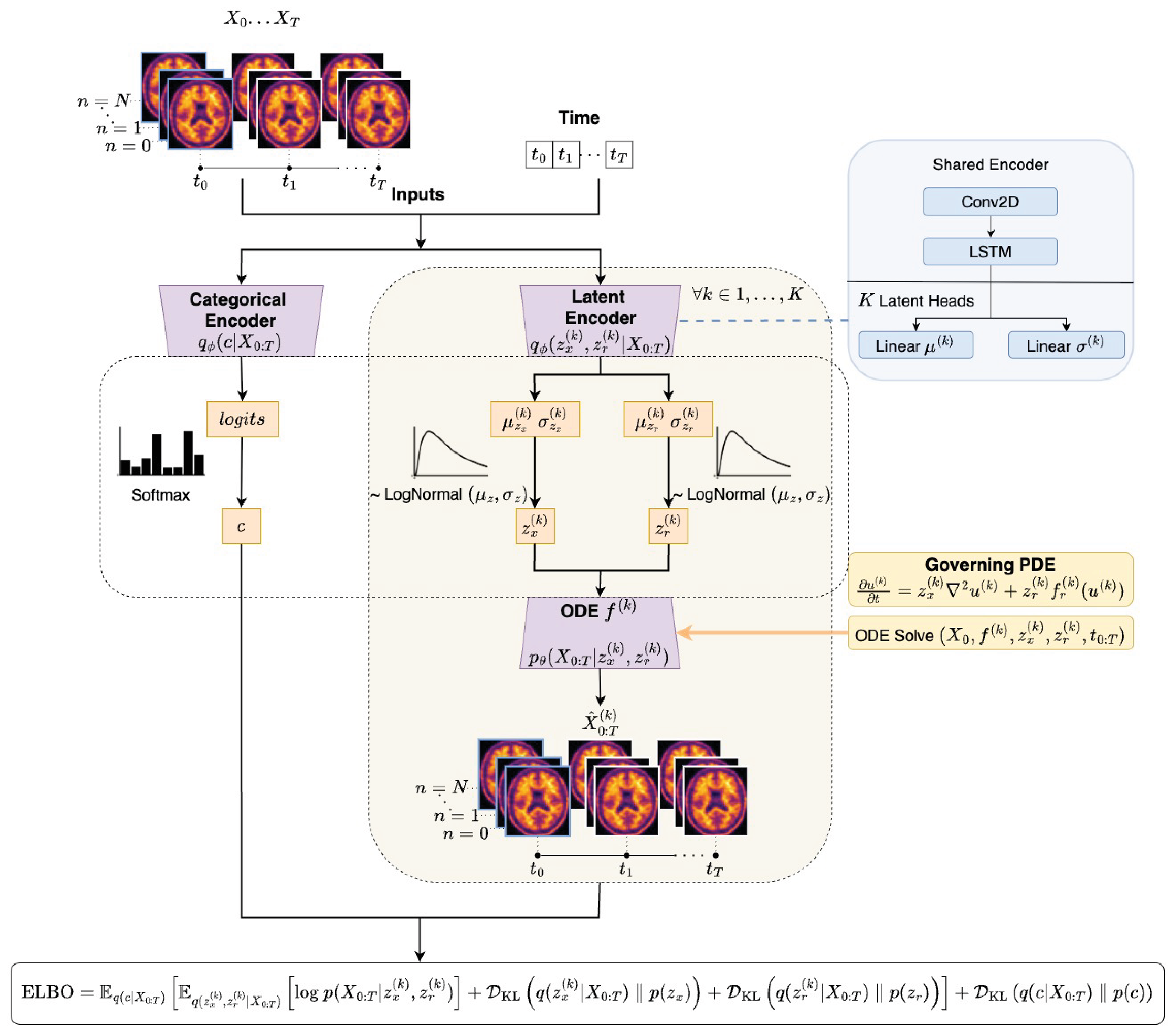}
\caption{\textbf{Schematic of the BrainPhys model.} 
Given a sequence of observations, a categorical encoder infers the mixture weights $c$, representing the probability of each candidate model being responsible for the observed dynamics. For each of the K candidate PDE models, a latent encoder infers the corresponding physical parameters $z_x^{(k)}$, $z_r^{(k)}$. Each component simulates disease dynamics using a differentiable ODE solver based on the component-specific KPP-type equation with the inferred parameters. The likelihood of the observed data is evaluated under each model, and the total data likelihood is computed as a weighted sum of these component likelihoods using the inferred mixture weights. This aggregate likelihood is used to compute and minimise the negative evidence lower bound (ELBO).}
\label{fig1}
\end{figure}

\normalsize
A schematic overview of our physics-informed VAE mixture model (``BrainPhys'') is given in Figure~\ref{fig1}. The source code is available at \footnote{\url{https://github.com/sanpinnawala/BrainPhys}}.
Expanding Equation~\ref{eq1} we define the diffusion coefficient as $D = z_x$ and the source term as $f(u) = z_rf_r(u)$,
\begin{equation}
\frac{\partial }{\partial t}u^{(k)} = z^{(k)}_x \nabla^2 u^{(k)} + z^{(k)}_r f^{(k)}_r(u^{(k)}).
\label{eq4}
\end{equation}
Here, we adopt a KPP-type model \cite{adomianFisherKolmogorovEquation1995} to describe the accumulation and propagation of misfolded protein. In our mixture of $k={1,2,3,...K}$ components, each component corresponds to a distinct formulation of the reaction term, $f^{(k)}_r$, with its own coefficients $z^{(k)}_x$,$z^{(k)}_r$, where the diffusion coefficient $z^{(k)}_x$ describes protein propagation, and the reaction coefficient $z^{(k)}_r$ describes protein accumulation, i.e., we make the form of the diffusion term common to all mixture components but allow the reaction term to vary. This decision was motivated by the hypothesis that the protein accumulation dynamics are more variable between individuals than the diffusion process.
\subsubsection{\normalsize Encoders} Each encoder branch, categorical and latent (Figure~\ref{fig1}), maps observed image sequences to a structured, low-dimensional latent space. Each image is passed through a convolutional backbone; the resulting features are concatenated with timestamps and passed to a long short-term memory (LSTM) module. The final LSTM hidden state is passed through task-specific fully connected layers.
\paragraph{Categorical Encoder.} The categorical encoder outputs a vector of logits representing unnormalised mixture weights over the $K$ components. We apply the \textit{softmax} function to obtain a probability vector $\mathbf{c} \in \Delta^{K-1}$. 
At inference time, we select the most probable component \(\hat{k} = \arg\max_k \, c_k.\)
\paragraph{Latent Encoder}
For each component $k \in \{1, \ldots, K\}$, the specific latent encoder outputs the mean and log variance for the physical parameters $(z_x^{(k)}, z_r^{(k)})$. To ensure positive values, physical parameters are sampled from log-normal distributions, where \(\tilde{z} = \exp{(z)}\) and \(z \sim \mathcal{N}(\mu,\sigma^2).\) $z$ is reparameterised using the reparameterisation trick \cite{kingmaAutoEncodingVariationalBayes2022}.
\subsubsection{\normalsize Decoder} 
Given the sampled physical parameters $\mathbf{z}^{(k)}$ and initial condition $X_0$, we use the differentiable ordinary/partial differential equation solver (\texttt{odeint}) from \texttt{torchdiffeq} \cite{chenTorchdiffeq2021} to integrate the PDE (discretised in space; see Equation~\ref{eq4}) forward in time. The solver outputs $\hat{X}^{(k)}$, the predicted dynamics from component $k$:
\begin{equation}
\hat{X}^{(k)} = \texttt{ODEINT}(X_0, f^{(k)}(z_x^{(k)}, z_r^{(k)}), \{t_0, \ldots, t_T\})
\label{eq5}
\end{equation}
\subsubsection{\normalsize Evidence Lower Bound}
The training objective is adapted to account for the component mixture. This is achieved by maximising the evidence lower bound (ELBO), incorporating both the reconstruction likelihood and latent regularisation.
The expected log-likelihood under the mixture model is computed using a log-sum-exp over the component-wise negative log-likelihoods:
\begin{equation}
\mathcal{L}_{\text{recon}} = \sum_{i=1}^N - \log \sum_{k=1}^{K} c_k^{(i)} \exp \left( - \mathcal{L}_{\text{NLL}}^{(k, i)} \right)
\label{eq6}
\end{equation}
where $c_k^{(i)}$ denotes the relaxed mixture weight for component $k$ and sample $i$, and $\mathcal{L}_{\text{NLL}}^{(k,i)}$ is the Gaussian negative log-likelihood (see Equation~\ref{eq12} in the Appendix).
$\mathcal{L}_{\text{KL}(z)}$ account for the divergence between the approximate and prior distributions over physical parameters, weighted by the mixture assignment $c_k^{(i)}$:
\begin{equation}
\mathcal{L}_{\text{KL}(z)} = \sum_i \sum_k c_k^{(i)} \left[ \text{KL}\left(q(z_x^{(i,k)} \mid X_i) \parallel p(z_x)\right) + \text{KL}\left(q(z_r^{(i,k)} \mid X_i) \parallel p(z_r)\right) \right]
\label{eq7}
\end{equation}
$\mathcal{L}_{\text{KL}(c)}$ is the KL divergence between the inferred categorical distribution $q(c \mid X)$ and the uniform prior over components $p(c) = \text{Cat}\left(\frac{1}{K}\right)$:
\begin{equation}
\mathcal{L}_{\text{KL}(c)} = \text{KL}\left(q(c \mid X) \parallel p(c)\right)
\label{eq8}
\end{equation}

\section{\large Experiments}
\subsection{\normalsize Synthetic Data}
\normalsize
We generate the synthetic dataset by numerically solving discretised Equation~\ref{eq1} using Runge-Kutta method of order 5 (RK45) with $\texttt{SciPy}$'s $\texttt{solve\_ivp}$ function. We define three distinct formulations of the reaction term, referred to as ID 0 (Equation~\ref{eq9}), ID 1 (Equation~\ref{eq10}), and ID 2 (Equation~\ref{eq11}), corresponding to three distinct clusters. The system is simulated on a \(32\times 32\) 2D spatial grid. We model initial conditions representing the initial protein concentration with Gaussian-like distributions over the spatial grid. We impose Neumann boundary conditions to ensure no flux across the domain boundaries. For each generated sample, we draw spatial and temporal parameters from uniform distributions: $z_x \in[0.01, 1.0]$ and $z_r\in[0.01, 0.1]$. Data normalisation scales values to the range $[0, 1]$. The training set consists of data samples representing 800 individuals with 3 observations per individual. The validation set consists of 200 data samples. For evaluation we generate a test set of 1000 samples.
\begin{equation}
\textbf{ID 0:}  \quad f_r(u) = z_ru(1-u) ~
\label{eq9}
\end{equation}
\begin{equation}
\textbf{ID 1:}  \quad f_r(u) = z_ru(1-u)^2
\label{eq10}
\end{equation}
\begin{equation}
\textbf{ID 2:}  \quad f_r(u) = z_ru^2(1-u)
\label{eq11}
\end{equation}

\subsection{\normalsize Alzheimer's Disease Data}
\normalsize
For Alzheimer's disease data, we use AV145-PET scans and AV45-PET scans from the Alzheimer’s Disease Neuroimaging Initiative (ADNI) \cite{petersenAlzheimerDiseaseNeuroimaging2010}. Access to the dataset is granted upon request through \cite{petersenAlzheimerDiseaseNeuroimaging2010}. We combine the two data types (tau and amyloid) comprising AD (Alzheimer's disease), MCI (mild cognitive impairment), and CN (cognitively normal) individuals. We hypothesise that there should be evidence for more than one cluster in this dataset. The ADNI data has been preprocessed for longitudinal consistency, including co-registration to the first time point for spatial alignment, spatial normalisation using SPM, and resampling to a uniform resolution of 6mm³. We apply intensity normalisation across dataset, scaling values to the range $[0,1]$. The training and validation sets consist of 237 and 60 individuals, respectively, each with 3 observations acquired at irregular time points.  

\section{\large Results}
\subsection{\normalsize Synthetic Data}
\normalsize
\begin{figure}[h!]
\centering
\includegraphics[width=\textwidth]{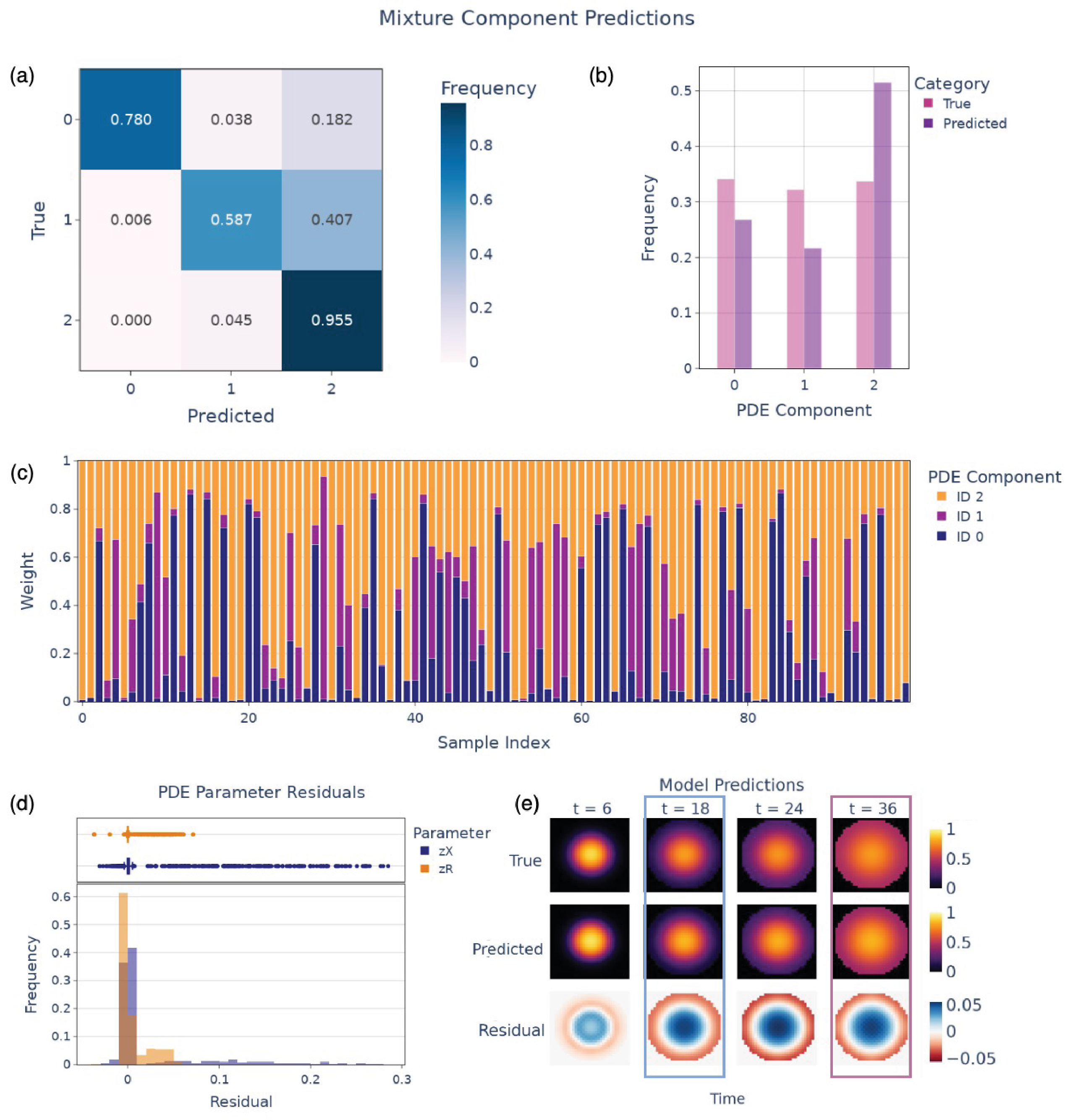}
\caption{\textbf{PDE component inference, PDE parameter inference, and predicted dynamics.} \textbf{(a)} Confusion matrix comparing the true cluster assignments (IDs 0, 1, 2) with the predicted assignments. \textbf{(b)} True vs. predicted distributions of component assignments. \textbf{(c)} Predicted mixture weights across 100 test samples. \textbf{(d)} residual distribution for the inferred diffusion and reaction coefficients. \textbf{(e)} Model predictions of a representative test sample.}
\label{fig2}
\end{figure}
\begin{figure}[h!]
\centering
\includegraphics[width=\textwidth]{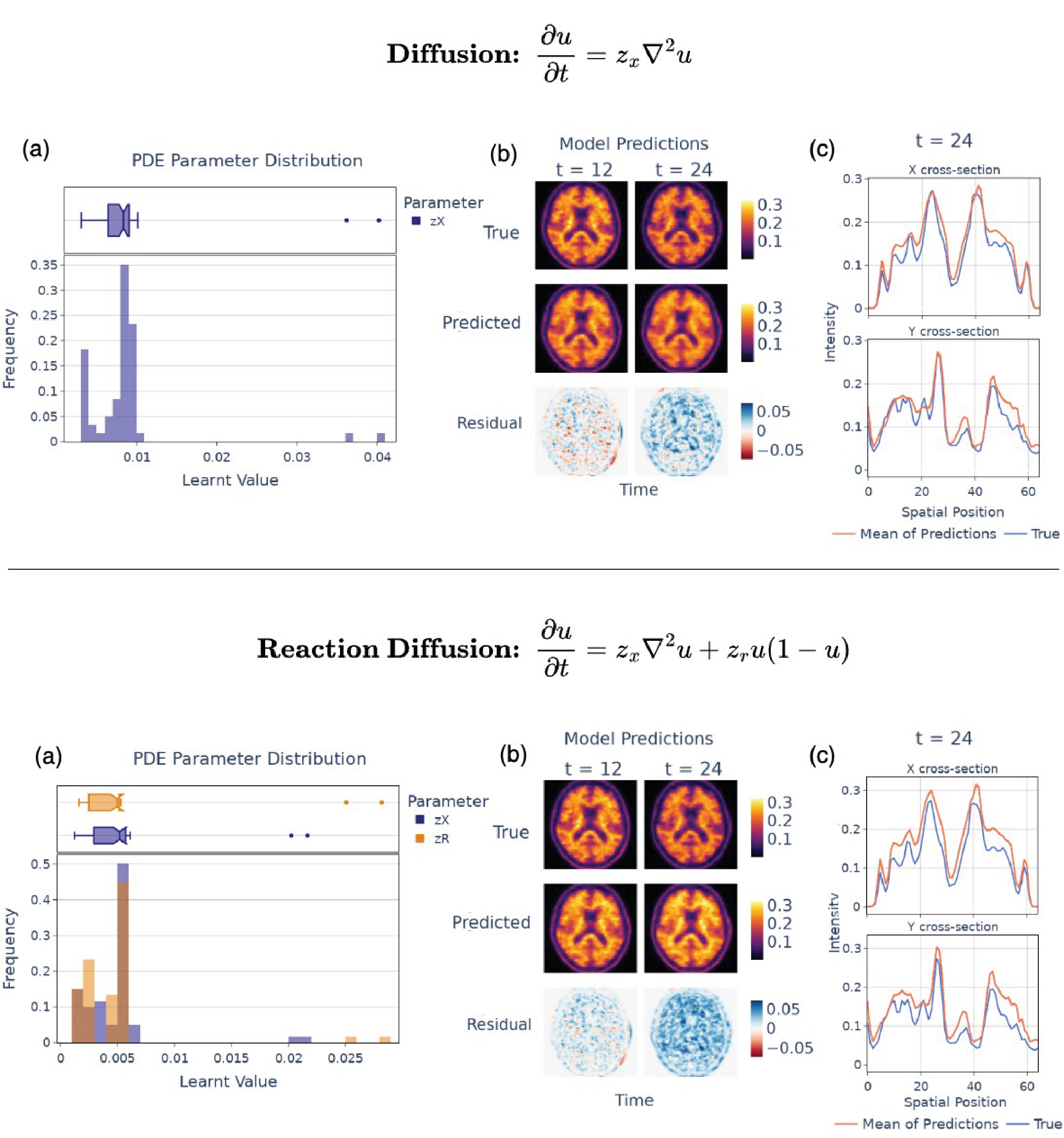}
\caption{\textbf{Evaluation on ADNI data using a diffusion and reaction-diffusion mixture model.} \textbf{(a)} Learnt parameter distribution. \textbf{(b)} Comparison of the mean prediction over 100 inferred parameter samples to the observations. Here, \(t\) refers to time in months from the baseline scan. \textbf{(c)} Central horizontal and vertical cross sections at the final reconstructed time point compared to the ground truth.}
\label{fig3}
\end{figure}
Figure \ref{fig2} presents results from the synthetic data experiment. Figure \ref{fig2}(a) and Figure \ref{fig2}(b) compare component assignments to the ground truth. Figure \ref{fig2}(c) shows the infered mixture weights for each PDE component across test samples, representing the model's belief about which component best explains each test sample. The assigned cluster corresponds to the component with the highest weight (see \textit{Categorical Encoder} part of Section~\ref{sec:2.2}). Figure \ref{fig2}(d) shows the residual distribution, computed as the difference between ground truth and predicted values from the assigned PDE component. The figure demonstrates successful inference of true parameters, notably the reaction rate. Figure \ref{fig2}(e) compares the ground truth to the mean prediction taken over 100 inferred parameter samples. Here, t=18 is an interpolated time point, and t=36 is an extrapolated one, demonstrating the model’s ability to generalise both within and beyond the observed data range. Model evidence across multiple runs is shown in Appendix Figure \ref{fig5}, reflecting variability attributable to random initialisation.

\subsection{\normalsize Alzheimer's Disease Data}
\normalsize
Figure \ref{fig3} describes results from the ADNI experiment using a two-component mixture model, where the components correspond to diffusion and reaction-diffusion PDEs. These PDEs reflect common competing hypotheses for tau and amyloid pathology. Since our model estimates the posterior over both components, we can infer the dynamics and parameters for each subject under either PDE. We compare the parameter distributions across the validation set and show reconstructions for a representative subject under both components. We evaluate the model evidence by recomputing the ELBO for each subset of mixture components at inference time. The evidence is highest when both PDE components are used \((2.07\times10^6)\), compared to using only reaction-diffusion \((1.96\times10^6)\), or only diffusion \((2.03\times10^6)\). The results suggest distinct clusters within the cohort, with the model benefiting from access to multiple PDE components.

\section{\large Discussion}
\normalsize
Here, we introduced ``BrainPhys'', a novel unsupervised methodology for inferring mixtures of mechanistic models from the same population. We demonstrated its ability to recover the parameters of PDE models and their respective weights from synthetic data. We also provided a first application of the model in ADNI, where we found evidence of a mixture of two components in a cohort of tau and amyloid PET data; however, further analysis is necessary to validate the clusters. Future work will extend this analysis to larger datasets. Our model also has potential for identifying sources of model degeneracy (see Figure \ref{fig4} in the Appendix), a problem that the mechanistic modelling community has yet to address. However, several challenges need to be addressed, including (but not limited to) disentangling factors contributing to ill-posedness, identifying model degeneracy when there is only one true model in the data, incorporating partial knowledge of PDE forms (e.g., unknown reaction terms), and inferring clusters within imaging modalities such as tau or amyloid PET alone.
\appendix
\section{\large Appendix}
\subsection{\normalsize Supplementary Methods}
\subsubsection{\normalsize Gaussian Negative Log-Likelihood}
\normalsize
\begin{equation}
\mathcal{L}_{\text{NLL}}^{(k,i)} = \frac{1}{2} \left[ \frac{\| X_i - \hat{X}_i^{(k)} \|^2}{\sigma^2} + D \log(2\pi \sigma^2) \right]
\label{eq12}
\end{equation}
where $D$ is the dimensionality of $X_i$ and $\sigma^2$ is the learned observation noise variance. KL term regularises the variational posteriors against its corresponding priors. 
\subsection{\normalsize Supplementary Results}
\begin{figure}[H]
\centering
\includegraphics[width=0.91\textwidth]{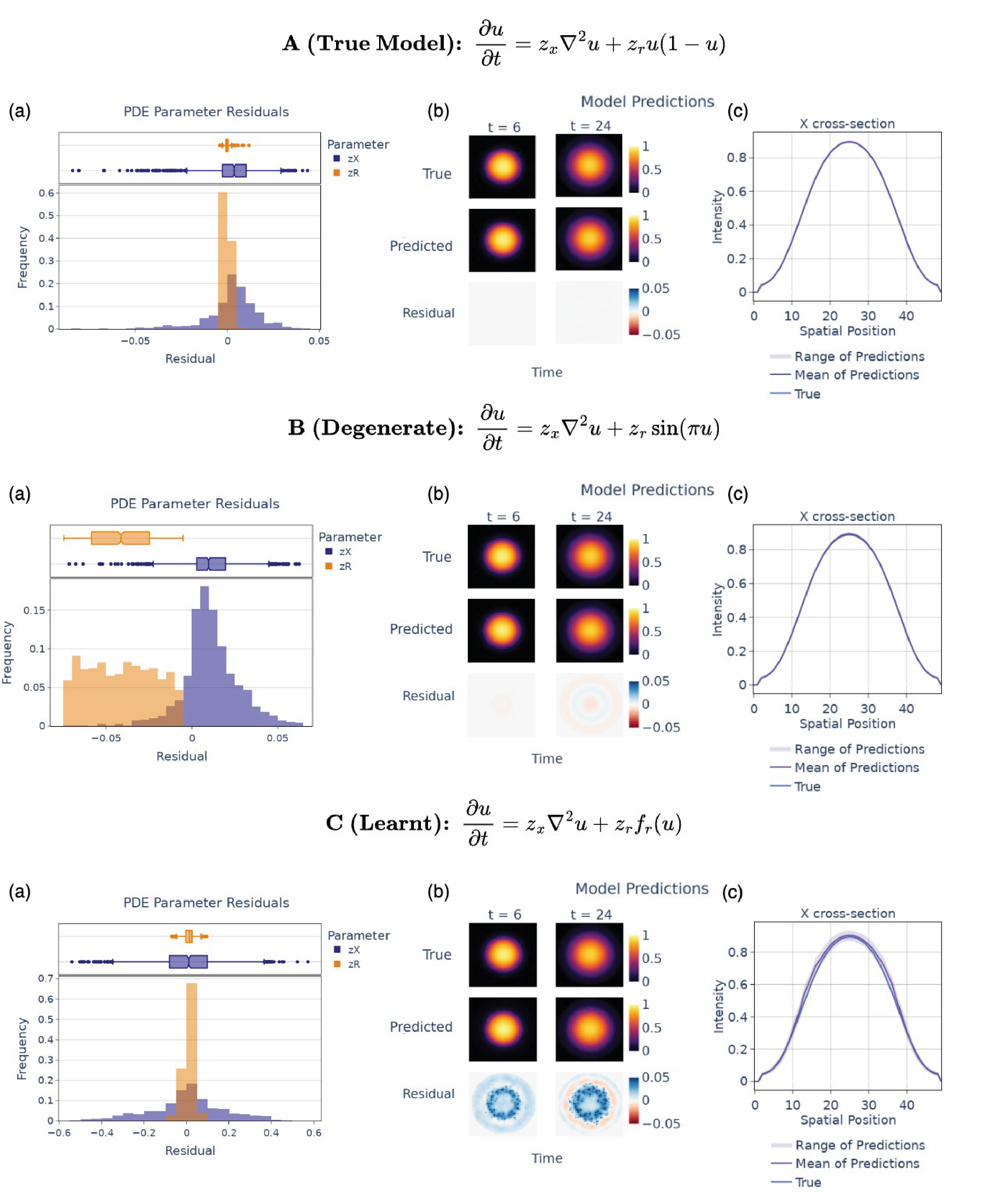}
\caption{\textbf{Testing model degeneracy on synthetic data.} \textbf{A.} Predictions from the true physics-based model used to generate the data. \textbf{B.} A degenerate model that achieves similar predictive accuracy but with inaccurate reaction rates, highlighting potential identifiability issues. \textbf{C.} The reaction term is treated as unknown and learnt using a neural network (MLP), demonstrating the model’s ability to approximate dynamics in the absence of explicit physical structure.}
\label{fig4}
\end{figure}
\begin{figure}[h!]
\centering
\includegraphics[width=\textwidth]{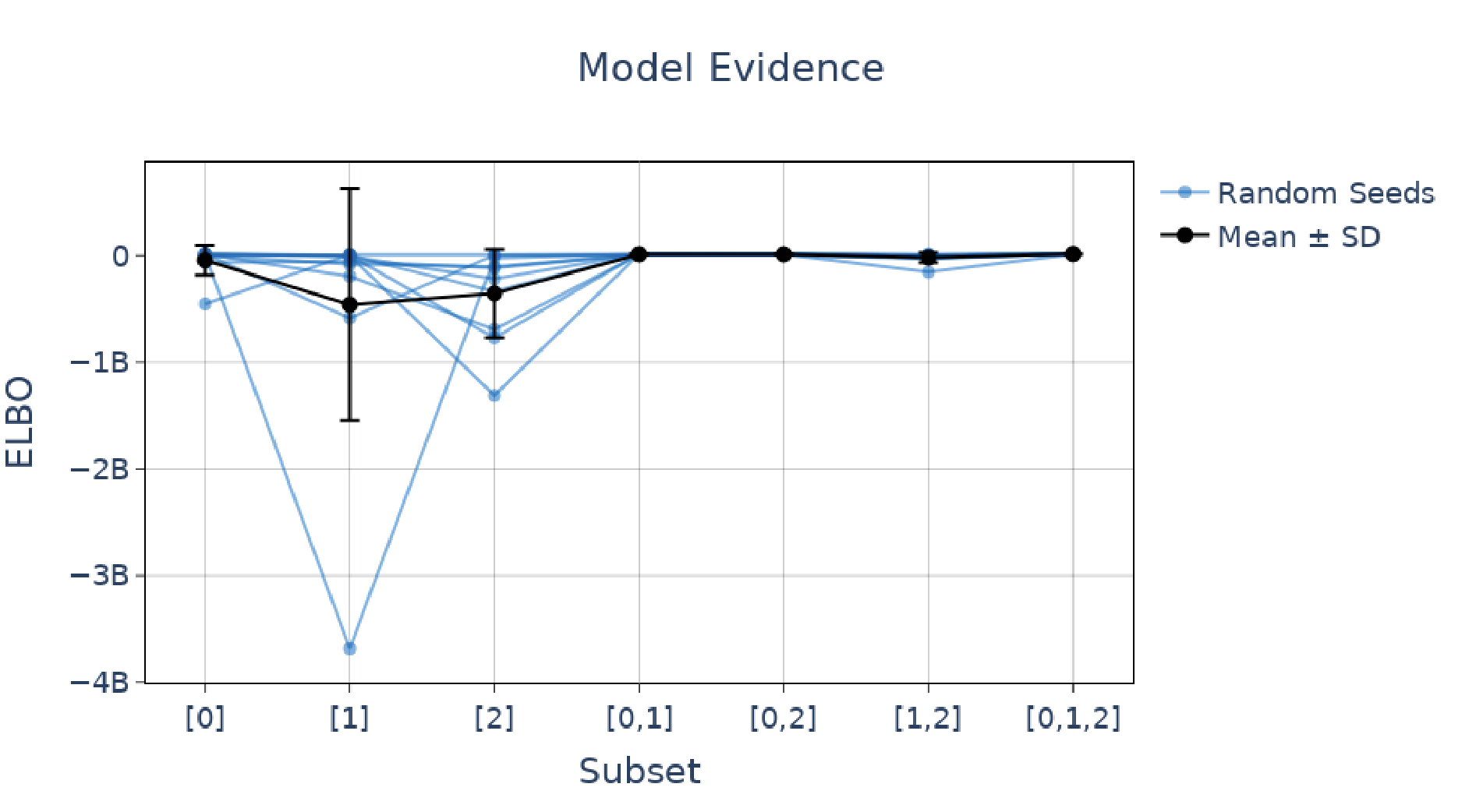}
\caption{\textbf{Model evidence for component subsets across runs.} Model evidence (i.e., ELBO) computed for each subset of mixture components across 10 random seeds in the synthetic data experiment (Section~3.1). The subset IDs correspond to the component definitions in Equation~\ref{eq9}--\ref{eq11}.}
\label{fig5}
\end{figure}

\bibliography{refs}
\end{document}